\documentclass[%
    reprint,
    superscriptaddress,
    a4paper,
    nofootinbib,
    amsmath,amssymb,
    aps,
]{revtex4-2}

\usepackage{graphicx} 
\usepackage{physics}

\usepackage[dvipsnames]{xcolor}
\usepackage[
    colorlinks,
    citecolor=RoyalBlue,
    linkcolor=RoyalBlue,
    urlcolor=black,
]{hyperref}

\newcommand{\iu}{\mathrm{i}\mkern1mu}

\newcommand{\affilANUnlpc}{Nonlinear Physics Center, Research School of Physics, Australia National University, Canberra ACT 2601, Australia}
\newcommand{\affilANUengineering}{School of Engineering, College of Science and Computer Science, Australian National University, Canberra, ACT 2601, Australia}

\begin{document}

\author{Pavel Tonkaev}
\email{pavel.tonkaev@anu.edu.au}
\thanks{equal contribution}
\affiliation{\affilANUnlpc}

\author{Ivan Toftul}
\thanks{equal contribution}
\affiliation{\affilANUnlpc}

\author{Zhuoyuan Lu}
\thanks{equal contribution}
\affiliation{\affilANUengineering}

\author{Shuyao Qiu}
\affiliation{\affilANUengineering}

\author{Hao Qin}
\affiliation{\affilANUengineering}

\author{Wenkai Yang}
\affiliation{\affilANUengineering}

\author{Kirill Koshelev}
\affiliation{\affilANUnlpc}

\author{Yuerui Lu}
\affiliation{\affilANUengineering}

\author{Yuri Kivshar}
\email{yuri.kivshar@anu.edu.au}
\affiliation{\affilANUnlpc}

\title{Nonlinear chiral metasurfaces based on structured van der Waals materials }

\begin{abstract}
Nonlinear chiral photonics explores nonlinear response of chiral structures, and it offers a pathway to novel optical functionalities not accessible through linear or achiral systems.  Here we present the first application of nanostructured van der Waals materials to nonlinear chiral photonics. We demonstrate the three orders of magnitude enhancement of the third-harmonic generation from hBN metasurfaces driven by quasi-bound states in the continuum and accompanied by strong nonlinear circular dichroism at the resonances. This novel platform for chiral metaphotonics can be employed for achieving large circular dichroism combined with high-efficiency harmonic generation in a broad frequency range.
\end{abstract}

\maketitle

\section{Introduction}

In the growing field of nanophotonics research of new materials that enhance light-matter interactions at the nanoscale is critical to developing modern photonic technologies. Hexagonal boron nitride (hBN), a two-dimensional van der Waals material, has proven to be a particularly promising candidate due to its distinctive properties~\cite{grudinin2023hexagonal}, including the large bandgap of hBN that ensures negligible absorption in the visible spectrum~\cite{cassabois2016hexagonal}.
Additionally, the van der Waals nature of hBN facilitates mechanical exfoliation and the manipulation of its layers, thereby enabling the construction of complex nanostructures optimized for light interaction. The ability of hBN to host quantum emitters makes it an important material for quantum photonic applications~\cite{froch2021coupling,li2021integration,spencer2023monolithic} enhanced by resonances~\cite{froch2022purcell,sortino2024optically}. Recent advancements in nanofabrication techniques have significantly enhanced the ability to precisely engineer hBN at the nanoscale and fabricate resonant nanostructures supporting Mie resonances~\cite{ling2024near}, high-$Q$ resonances~\cite{das2021demonstration} and bound states in the continuum (BIC)~\cite{gupta2023bound,kuhner2023high}. Additionally, hBN structures exhibit strong third-harmonic generation (THG) response~\cite{bernhardt2021large,popkova2021optical}. These factors are instrumental in realizing the full potential of hBN in nanophotonic applications, leading to the creation of metasurfaces capable of manipulating chiral nonlinear optical fields with high efficiency.

Chiroptical effects can be characterized by {\it circular dichroism} (CD)~\cite{berova2000circular}, which involves the asymmetric transmission or absorption of circularly polarized light in a chiral medium.  In natural materials, optical activity is generally very weak. However, artificially engineered materials with macroscopic chirality have been developed to enhance these weak natural chiroptical effects over the past two decades. Starting from three-dimensional chiral metamaterials were shown to enhance CD in the linear regime~\cite{pendry2004chiral,wang2009chiral, oh2015chiral},  the research focus has shifted towards the study of planar chiral structures composed of ordered arrays of subwavelength elements which were demonstrated to change the polarization state of light in a way similar to three-dimensional chiral media but require simpler fabrication technologies~\cite{papakostas2003optical,kuwata2005giant,konishi2008observation, volkov2009optical}. The efficiency of linear chiroptical response can further be improved by employing resonant metaphotonic structures made of plasmonic and dielectric materials~\cite{valev2013chirality, hentschel2017chiral}. This laid the foundation of new applications for chiral emission~\cite{zhang2022chiral}, chiral sensing~\cite{mohammadi2018nanophotonic, solomon2018enantiospecific}, characterization of quantum light~\cite{wang2023characterization}, and chiral harmonic generation~\cite{koshelev2023nonlinear}.

Natural nonlinear optical activity is significantly more pronounced than linear chiroptical effects because the optical harmonics generated by chiral light are highly sensitive to molecular and structural asymmetry~\cite{verbiest1999light}, which makes it a powerful tool for chiral biosensing~\cite{Petralli-Mallow1993JPhysChem}. That attracted special interest to  second- and third-harmonic CD of bulk media and planar interfaces~\cite{byers1994second,maki1995surface,verbiest1998strong}. Similar to linear effects, the enhancement of nonlinear chiroptical effects beyond the limit of natural material response can be achieved by engineering chirality at macroscopic scale~\cite{haupert2009chirality, li2017nonlinearR, konishi2020tunable, rodrigues2022review}. Unlike the linear regime, strong nonlinear optical activity also requires high nonlinear conversion efficiency. Therefore, the resonances of metaphotonic structures are expected to play a crucial role, as they have been shown to dramatically improve frequency conversion efficiency by enhancing light-matter interaction~\cite{shi2022planar,koshelev2023resonant}. Thus, nonlinear chiral photonics, in combination with new materials such as wan der Waals materials, may offer a pathway to novel optical phenomena and functionalities not accessible through linear or achiral systems.

\begin{figure*}
    \centering
    \includegraphics[width=0.7\linewidth]{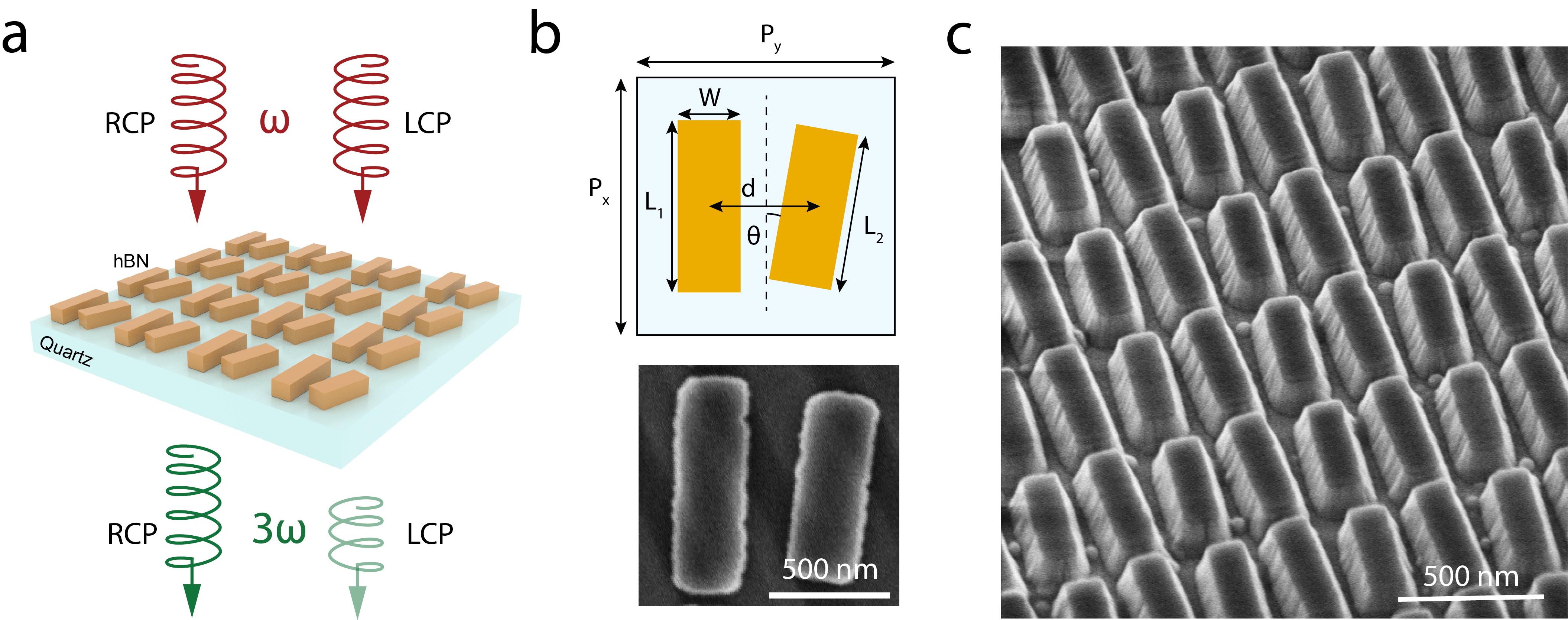}
    \caption{ \textbf{Chiral nonlinear metasurfaces}. (a) Schematic image of hBN metasurface for chiral third harmonic generation. (b) Schematic and SEM images of meta-atom and (c) SEM image of the metasurface.
    }
    \label{fig1}
\end{figure*}

In this paper, we study dielectric metasurfaces to enhance nonlinear response and present the first application of hBN-based metasurfaces in nonlinear chiral photonics, demonstrating the generation of chiral optical harmonics through nonlinear optical processes. We engineer and fabricate nonlinear hBN metasurfaces supporting quasi-BIC mode in the near-IR range. We predict theoretically and demonstrate experimentally three orders of magnitude enhancement of the third harmonic generation compared to the thin film driven by quasi-BIC. We show that nonlinear CD changes from the value +0.12 to -0.38 with pump wavelength. We believe the exploration of hBN in this capacity underscores its potential to contribute significantly to the development of advanced photonic devices and systems and our results pave the way toward a new platform for biosensing in the whole visible range and near UV based on nanostructured van der Waals materials.

\section{Results and Discussion}

Our metasurfaces consist of meta-atoms placed on top of fused silica substrate arranged into a square lattice, as shown in Figure~\ref{fig1}a. The unit cell is two bars, one of which is shorter in length and rotated on an angle relative to the axis between the bars. The geometrical parameters of our meta-atoms are $L_1 = 864$~nm, $L_2 = 744$~nm, $W = 240$~nm, $d = P_y/2 = 538$~nm, $\theta = 5.5^{\circ}$, $P_x = 1025$~nm, $P_y = 1075$~nm, as shown in Figure~\ref{fig1}b. We engineer the metasurface to support quasi-BIC mode in the near-infrared spectral range. We fabricate the metasurface from hBN film, and the scanning electron microscopy (SEM) image of the metasurface obtained is shown in Figure~\ref{fig1}c (for fabrication process details see Supporting Information S1).

\begin{figure*}
    \centering
    \includegraphics[width=0.7\linewidth]{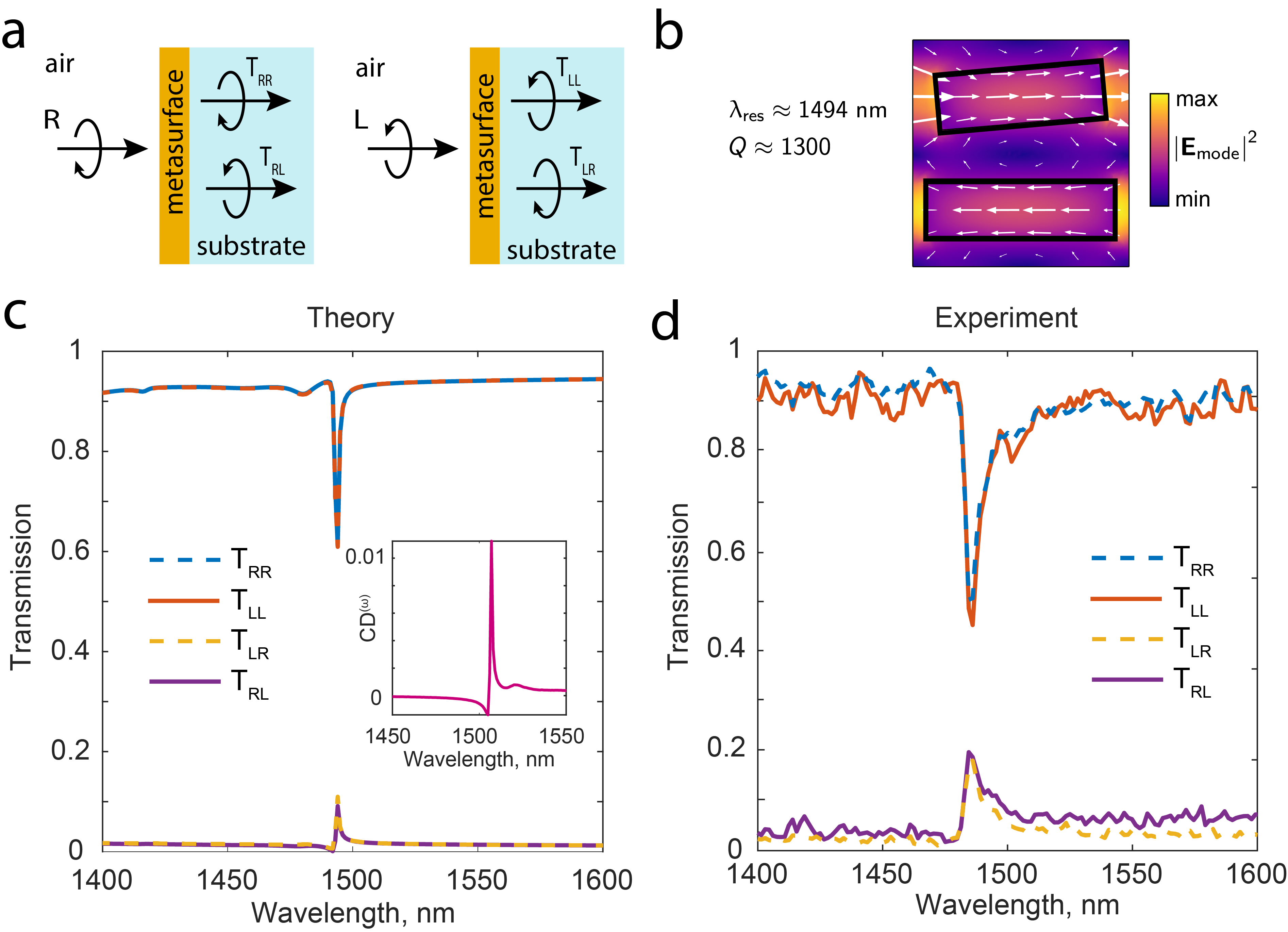}
    \caption{ \textbf{Linear Properties}. (a) Schematic image defining of different transmission parts.  (b) Electrical field distribution in meta-atom for the light with a wavelength of 1494 nm. Theoretical (c) and experimental (d) transmission of the metasurface for LCR and RCP light. The theoretical CD$^{(\omega)}$ is shown in the inset.}
    \label{fig2}
\end{figure*}

First, we characterise the linear chiroptical response of the metasurfaces by measuring co-polarised and cross-polarised transmission. To measure it we illuminate the metasurface normally with left-circularly polarized (LCP) and right-circularly polarized (RCP) near-infrared light. We use a quarter-wave plate in the collection to separate the passed LCP and RCP light. By normalisation of the passed light intensity on the incident light intensity with different polarization, we obtained four transmission coefficients corresponding to co-polarised light with RCP (T$_{RR}$) and LCP (T$_{LL}$), and cross-polarised light with RCP in pump and LCP in the collection (T$_{RL}$) and LCP in pump and RCP in the collection (T$_{LR}$) (Figure~\ref{fig2}a).

The numerical simulations reveal that transmission of the metasurface exhibits resonant behaviour at the wavelength of 1494 nm, which is observed as a dip for co-polarised light and a peak for cross-polarised light (Figure~\ref{fig2}c). This resonant dip for co-polarised light is caused by the excitation of quasi-BIC mode. The electric field distribution of the mode supported is shown in Figure~\ref{fig2}b. The electric field vectors in different bars are oppositely directed.

The transmission for co-polarised light has a small difference for RCP and LCP. To determine the linear CD we use the following definition: 
\begin{equation}
\mathrm{CD}^{(\omega)} = \frac{T_{\text{LL}} - T_{\text{RR}}}{T_{\text{LL}} + T_{\text{RR}}}
\end{equation} According to numerical calculations, the linear CD is typically around +0.01 at the resonance, and it vanishes for the wavelengths outside the resonance,  as shown in the inset of Figure~\ref{fig2}c.  

The experimentally obtained transmission spectra for co-polarized and cross-polarized light are shown in Figure~\ref{fig2}d (for more details of the experiment see Supporting Information S2). The spectra for RCP and LCP co-polarised light demonstrate the dip at the wavelength of 1494 nm and no significant difference between them. Meanwhile, the cross-polarized transmission spectra show a value different from the zero only in the vicinity of the resonance. The behaviour observed is in good agreement with theoretical prediction.

\begin{figure*}
    \centering
    \includegraphics[width=0.7\linewidth]{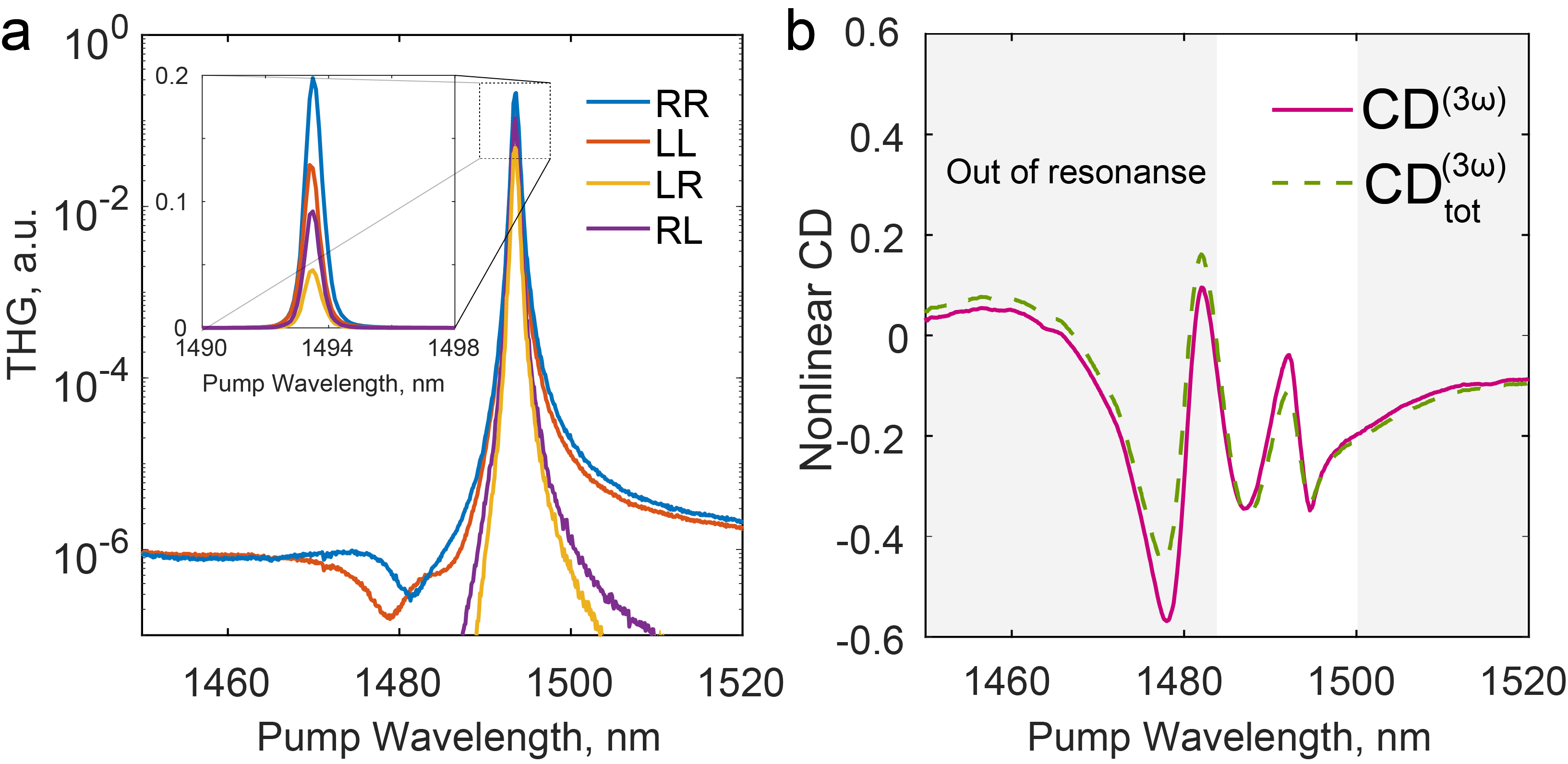}
    \caption{ \textbf{Nonlinear Properties}.  (a) Calculated third harmonic generation for co-polarized RCP (blue) and LCP (red) excitation and collection signals; calculated RCP third harmonic signal excited by LCP light (yellow) and LCP third harmonic signal excited by RCP light (purple). (b) Calculated nonlinear CD of third harmonic for full signal (green) and co-polarized excitation and collection (pink).
    }
    \label{fig3}
\end{figure*}

Next, we simulate and measure the THG for the designed hBN metasurface and evaluate the nonlinear CD. For nonlinear simulation, we use the undepleted pump approximation. The THG signal is calculated in the transmission direction in the zero diffraction order (for more details see Supporting Information S3). The spectra obtained for co-polarised pump light and harmonic signal with RCP (RR) and LCP (LL) are shown in Figure~\ref{fig3}a by blue and red curves, respectively. We also acquired the RCP THG signal generated from the structure pumped by LCP (RL) and the LCP THG signal generated from the structure pumped by RCP (LR), as shown in Figure~\ref{fig3}a by purple and yellow curves, respectively. The simulated THG signal exhibits enhancement of six orders of magnitude at the pump wavelength of 1494 nm compared to the signal from the pump outside the resonance. The zoomed-in area near the resonance is shown in Figure~\ref{fig3}a inset. One can see that the amplitude of the THG signal depends on the pump polarization. To evaluate the nonlinear CD we use the two terms taking into account only co-polarised signal ($ \mathrm{CD}^{(3\omega)}$) and the full signal pumped ($\mathrm{CD}_{\text{tot}}^{(3\omega)}$) defined as

\begin{equation}
    \mathrm{CD}^{(3\omega)} = \frac{I_{\text{LL}}^{(3\omega)} - I_{\text{RR}}^{(3\omega)}}{I_{\text{LL}}^{(3\omega)} + I_{\text{RR}}^{(3\omega)}}
    \label{eq2}
\end{equation}

\begin{equation}
    \mathrm{CD}_{\text{tot}}^{(3\omega)}= \frac{I_{\text{L}}^{(3\omega)} - I_{\text{R}}^{(3\omega)}}{I_{\text{L}}^{(3\omega)} + I_{\text{R}}^{(3\omega)}},
    \label{eq3}
\end{equation}
where $I_{\text{LL(RR)}}^{(3\omega)}$ is the intensity of the forward generated LCP(RCP) THG signal with LCP(RCP) pump polarization, $I_{\text{L(R)}}^{(3\omega)}$  is the full forward intensity of THG signal with LCP(RCP) pump polarization.
The resulting nonlinear CD curves are shown in Figure~\ref{fig3}b. Both curves exhibit a similar trend. The maximum of nonlinear CD value is observed at the wavelength of 1475 nm. However, this value corresponds to a low THG signal outside the resonance, and at the vicinity of the resonance, the nonlinear CD value changes from -0.35 to 0. 

\begin{figure*}
    \centering
    \includegraphics[width=0.99\linewidth]{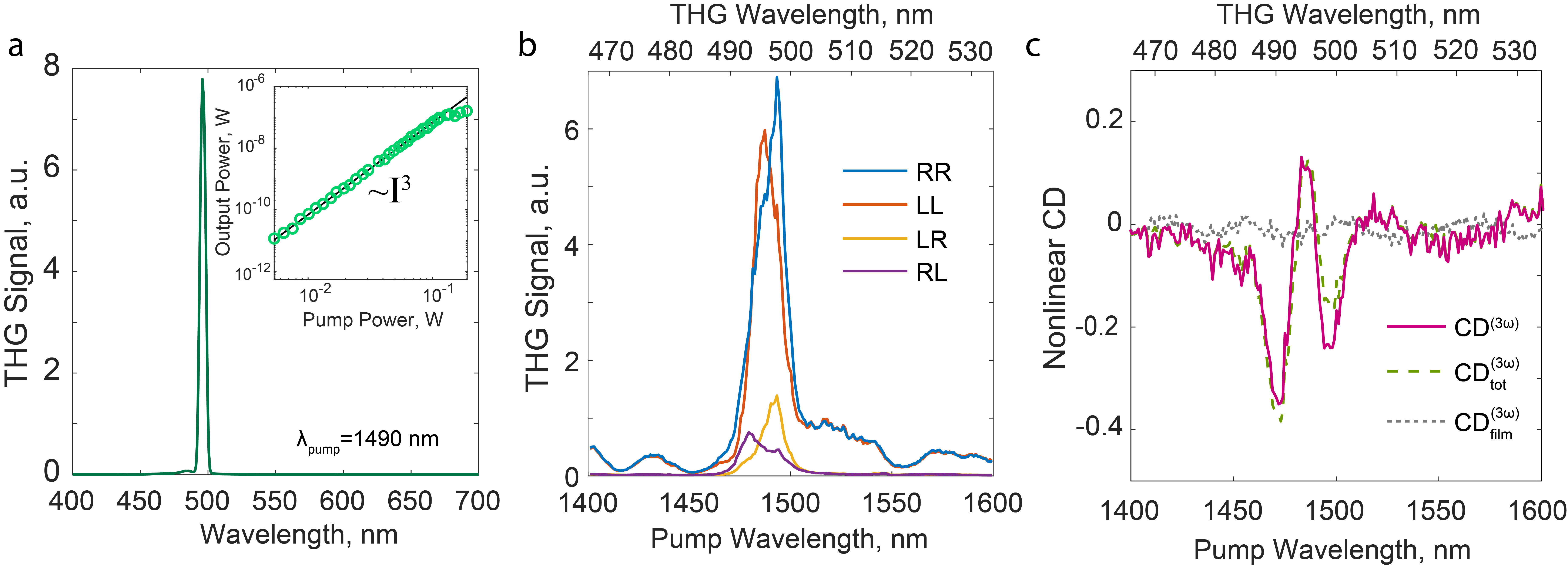}
    \caption{ \textbf{Chiral third-harmonic generation}. (a) A spectrum of the third harmonic from the metasurface pumped at the wavelength of 1490 nm. The inset shows the power-power curve of the third harmonic. (b) The third harmonic signal co-polarized RCP (blue) and LCP (red) excitations and collection, and for cross-polarised signal excited by LCP (yellow) and RCP (purple) pump. (c) Nonlinear CD of third harmonic for full signal (green) and co-polarized excitation and collection (pink). The grey line represents the nonlinear CD of the third harmonic from the unpatterned film. 
    }
    \label{fig4}
\end{figure*}

We further measure THG from fabricated hBN metasurfaces pumped in the near-IR range from 1400 nm to 1600 nm (for more details see Supporting Information S2). The spectrum of the THG pumped at a wavelength of 1490 nm has a maximum at a wavelength of 497 nm and is shown in Figure~\ref{fig4}a. The power dependence recorded has a good agreement with cubic as shown in the inset of Figure~\ref{fig4}a. The maximum THG efficiency of 10$^{-6}$ is observed at the pump wavelength of 1490 nm and the peak intensity of 10~GW/cm$^2$. We notice that only zero diffraction order of THG is collected, so the full THG efficiency tends to be higher.

To investigate the nonlinear CD, we pump the metasurfaces with LCP and RCP and separate the LCP and RCP of THG in the collection. We pump the metasurface with wavelengths from 1400 nm to 1600 nm and record the THG spectra. Next, we determine the maximum of the THG for every pump wavelength. The maximum of the THG for different pump wavelengths is shown in Figure~\ref{fig4}b. The THG signal strongly depends on the pump wavelength. At the resonance, the THG signal is enhanced by about three orders of magnitude compared to the THG signal from the thin hBN film (see Supporting Information S4). Based on Eqs~\ref{eq2} and~\ref{eq3}, we calculate nonlinear CD for only co-polarised components (CD$^{(3\omega)}$) and total nonlinear CD without separation of LCP and RCP in the collection (CD$_{\text{tot}}^{(3\omega)}$). The resulting dependencies are shown in Figure~\ref{fig4}c. The total nonlinear CD without separation of LCP and RCP in the collection shows the switching of the value from -0.38 to +0.12 near the resonance of the structure, whereas the nonlinear CD of the only co-polarised components exhibits the change from -0.35 to +0.12 near the resonance. The difference between the numerically simulated spectra and the experimental results can be attributed to at least two factors: \textit{(1)} side walls of fabricated bars are not exactly vertical but slightly inclined (see Fig.~\ref{fig1}c); and \textit{(2)}  in numerical simulations we assumed that the crystallographic axes are aligned with the metasurface grating direction, which could be different in the real structure, and extremely hard to control.

In addition, we observe the second-harmonic generation (SHG) from the metasurfaces in the vicinity of the resonance. The SHG is several orders of magnitude lower than the THG and the power dependence demonstrates a good agreement with quadratic law. The SHG pumped at the resonance wavelength exhibits different amplitude for LCP and RCP pumps. To estimate the nonlinear CD for SHG,  we used the similar definitions, $\mathrm{CD}^{(2\omega)} = (I_{\text{LL}}^{(2\omega)} - I_{\text{RR}}^{(2\omega)})/(I_{\text{LL}}^{(2\omega)} + I_{\text{RR}}^{(2\omega)})$ and $\mathrm{CD}_{\text{tot}}^{(2\omega)} = (I_{\text{L}}^{(2\omega)} - I_{\text{R}}^{(2\omega)})/(I_{\text{L}}^{(2\omega)} + I_{\text{R}}^{(2\omega)})$, where $I_{\text{LL(RR)}}^{(2\omega)}$ is the intensity of the forward generated LCP(RCP) SHG signal with LCP(RCP) pump polarization, $I_{\text{L(R)}}^{(2\omega)}$  is the full forward intensity of THG signal with LCP(RCP) pump polarization. The total second harmonic CD$_{\text{tot}}^{(2\omega)}$ shows the value of -0.27 at the resonance of the structure, whereas for co-polarised collection CD$^{(2\omega)}=+0.24$ (for more details, see Supporting Information S5).

\section{Conclusions}

In conclusion, we have observed the three orders of magnitude enhancement of the third harmonic generation from hBN metasurfaces driven by quasi-BIC mode. We have experimentally demonstrated strong nonlinear chiral response of the metasurface at the resonances. We believe our results introduce a new platform for chiral photonics, based on nanostructured van der Waals materials supporting optical resonances, which can be employed for chiral biosensing in the whole visible and near UV frequency ranges.

\begin{acknowledgements}
This work has been supported by the Australian Research Council (grant No. DP210101292) and the International Technology Center Indo-Pacific (ITC IPAC) via Army Research Office (contract FA520923C0023). Y. Lu acknowledges a funding from the Australian Research Council (grants Nos. DP240101011, DP220102219, and LE230100113). The authors acknowledge the use of the facilities at the ACT Node of the NCRIS-enabled Australian National Fabrication Facility at the Australian National University (ANFF-ACT).
\end{acknowledgements}

\appendix
\section{Methods}

\subsection{Samples Fabrication}

hBN thin film is mechanically exfoliated from bulk crystal (HQ graphene) with PDMS thin film and then transferred to a fused silica substrate. The thin film with targeted thickness is characterized by a Stylus profilometer (Bruker Dektak) and AFM. Then a layer of ZEP 520A resist with a thickness of 370 nm was spin-coated on the sample and baked at 160 degrees for 120 s. After that 10 nm Au was thermal deposited on the top of the resist to to avoid electron charge accumulation. The lithography pattern was defined using electron beam lithography (Raith 150) with an acceleration voltage of 20 kV, aperture size of 20 µm, a working distance of 10 mm, and an area dose of 75 $\mu$C/cm$^2$. After the patterning, the sample was dipped into the potassium-based gold etchant for 60s and the gold was removed thoroughly. The ZEP photoresist was dipped into the ZEP developer for 60s then IPA for 30s. 40 nm Ni was then deposited on the pattern. The sample is then left for lift-off for 2 hours. ICP-F (Samco 400iP) is then used to etch the hBN. The Ni mask is then washed with FeCl$_3$ solution.

\subsection{Optical Measurements}

The laser system consists of a 1030 nm laser (Ekspla Femtolux 3) and an optical parametric amplifier (MIROPA from Hotlight Systems). The laser has a pulse duration of $575$~fs and a repetition rate of $5.14$~MHz. The optical parametric amplifier produces near-IR radiation as signal pulses from the amplification of continuous wave spectrally narrow seed lasers in the near-IR spectrum. The resulting wavelength is tunable in the range of 1400-1700 nm. The laser was circularly polarised by the polariser and Thorlabs near-IR achromatic quarter waveplate. The near-IR radiation was focused with the CaF$_2$ lens with a focus of 40 mm on the sample. The harmonic signal was collected by  Mitutoyo MPlan Apo X50 0.4 NA microscope objective and detected with a Peltier-cooled spectrometer Ocean Optics QE Pro. For dividing the LCP and RCP harmonic signal the polariser and Thorlabs visible achromatic quarter waveplate were used. For transmission measurement, the same geometry was employed with the exchange of visible quarter waveplate in the collection by superachromatic quarter waveplate and near-IR spectrometer Ocean Optics NIR Quest.

\subsection{Numerical Simulations}

All numerical simulations were performed in the Wave Optics module of COMSOL Multiphysics.
The near-field distributions, resonant wavelengths, and $Q$ factors are simulated using the eigenmode solver.
Linear and nonlinear transmission simulations are simulated in the frequency domain.
Metasurface was placed on a semi-infinite substrate surrounded by a perfectly matched layer mimicking an infinite region in the vertical direction.
The simulation area is the unit cell with Floquet periodic boundary conditions which simulate an infinite size of the metasurface in a transverse plane.
The dispersion of the refractive index for hBN is taken from Ref.~\cite{grudinin2023hexagonal}, while for the SiO$_2$ is from Refs.~\cite{Malitson1965Oct,Polyanskiy2024Jan}.
The background field is set manually via custom code using Fresnel equations.
The third harmonic process is calculated in the undepleted pump approximation using the domain polarization feature~\cite{COMSOL_SHG_example}. The nonlinear polarization current is calculated as $P_{i}^{(3\omega)} = \varepsilon_0 \hat{\chi}^{(3)}_{ijkm} E_{j}^{(\omega)} E_{k}^{(\omega)} E_{m}^{(\omega)}$, where $\hat{\chi}^{(3)}$ tensor has 21 nonzero elements based on the hBN point symmetry group $\mathrm{D}_{6 \mathrm{h}}$ (196-th space group)~\cite{boyd2008nonlinear}. 
We assume that the crystallographic axes are aligned with the
metasurface grating direction and incident field direction, i.e. with the base Cartesian unit vectors $(\mathbf{\hat{x}}, \mathbf{\hat{y}}, \mathbf{\hat{z}})$.
Once the total fields are calculated for the RCP and LCP background fields, $\mathbf{E}_{^{\text{R}} _{\text{L}}}$, the  complex transmission amplitude coefficients  are calculated as $t^{(n\omega)}_{^{\text{RR}} _{\text{LL}} } = \braket{\mathbf{\hat{e}}_{\pm}}{\mathbf{E}_{^{\text{R}} _{\text{L}}}^{(n\omega)}} =  \frac{1}{A} \iint\limits_{A} \mathbf{\hat{e}}^{*}_{\pm} \cdot \mathbf{E}_{^{\text{R}} _{\text{L}}}^{(n\omega)} (x,y,z_0) \dd x \dd y$, where $A$ is the area of the $z=z_0$ plane located at the edge of the simulation area in the substrate, and $\mathbf{\hat{e}}_{\pm} = (\mathbf{\hat{x}} \pm \iu \mathbf{\hat{y}} )/\sqrt{2}$ are the unit vector in the circular basis.
Integration over surface $A$ averages the output signal over the angles, so it gives only the $0$-th diffraction order.
Finally, the transmission coefficients are calculated as $T^{(\omega)}_{^\text{RR} _{\text{LL}}} = \frac{n_{\text{subs}}}{n_{\text{host}}} \abs{t_{^\text{RR} _{\text{LL}}}^{(\omega)}}^2$, and the output nonlinear intensity is $I_{^\text{RR} _{\text{LL}}}^{(3\omega)} \propto \abs{t_{^\text{RR} _{\text{LL}}}^{(3\omega)}}^2$, where the proportionality coefficient is unimportant within the scope of this work.

\bibliography{refs}

\end{document}